\documentclass[review, authoryear]{elsarticle}
\usepackage{graphicx}
\usepackage[hidelinks]{hyperref}
\usepackage{epstopdf}
\usepackage{amsmath}
\usepackage{amsfonts}
\usepackage{url}
\usepackage{adjustbox}
\usepackage{textcomp}
\usepackage{siunitx,booktabs}

\usepackage{array}   
\newcolumntype{L}{>{$}l<{$}} 

\journal{Speech Communication}

\biboptions{authoryear}

\begin{document}

\begin{frontmatter}

\title{Two-stage dimensional emotion recognition by fusing predictions of acoustic and text networks using SVM}

\author[mymainaddress,mysecondaryaddress]{Bagus Tris Atmaja\corref{mycorrespondingauthor}}
\cortext[mycorrespondingauthor]{Corresponding author, \emph{E-mail address: bagus@ep.its.ac.id}}

\author[mymainaddress]{Masato Akagi}

\address[mymainaddress]{Japan Advanced Institute of Science and Technology, 
 1-1 Asahidai, Nomi, Ishikawa 923-1292, Japan}
\address[mysecondaryaddress]{Sepuluh Nopember Insitute of Technology, 
 Sukolilo, Surabaya 60111, Indonesia}

\begin{abstract}
Automatic speech emotion recognition (SER) by a computer is a critical
component for more natural human-machine interaction. As in human-human
interaction, the capability to perceive emotion correctly is essential to
take further steps in a particular situation. One issue in SER is whether it
is necessary to combine acoustic features with other data, such as facial
expressions, text, and motion capture. This research proposes to combine
acoustic and text information by applying a late-fusion approach consisting of
two steps. First, acoustic and text features are trained separately in deep
learning systems. Second, the prediction results from the deep learning systems
are fed into a support vector machine (SVM) to predict the final regression
score. Furthermore, the task in this research is dimensional emotion modeling,
because it can enable a deeper analysis of affective states. Experimental
results show that this two-stage, late-fusion approach obtains higher
performance than that of any one-stage processing, with a linear correlation
from one-stage to two-stage processing.  This late-fusion approach improves
previous early fusion results measured in concordance correlation coefficients
score.
\end{abstract}


\end{frontmatter}


\section{Introduction}
Understanding human emotion is important for responding properly in a
particular situation for both human-human communication and future
machine-human communication. Emotion can be recognized from many modalities:
facial expressions, speech, and motion of body parts. In the absence of visual
features, speech is the only way to recognize emotion, as in the case of a
telephone call or a call-center application \citep{Petrushin1998}. By
identifying caller emotions automatically from a system, appropriate feedback
can be applied quickly and precisely.

Speech is a modality in which both acoustic and verbal information can be
extracted to recognize human emotion. Unfortunately, most speech emotion
recognition (SER) systems use only acoustic features for predicting categorical
emotions. In contrast, this research proposes to use both acoustic and text
features to improve dimensional SER performance. Text can be extracted from
speech, and it may contribute to emotion recognition. For example, an
interlocutor can perceive emotion not only from prosodic information but also
from semantics. \cite{Grice2002} stated in his implicature theory that what is
implied derives from what is said. For example, if someone says that he is
angry but looks happy, then the implication is that he is indeed angry. Hence,
it is necessary to use linguistic information to determine
expressed emotion from speech. A fusion of acoustic and linguistic information
from speech is viable since (spoken) text can be obtained from speech-to-text
technology. This bimodal features fusion strategy may improve the performance
of SER over acoustic-only SER.

Besides the categorical approach, emotion can also be analyzed via a dimensional
approach. In dimensional emotion, affective states are lines in a continuous
space. Some researchers have used a two-dimensional (2D) space comprising
valance (positive or negative) and arousal (excited or apathetic). Other
researchers have proposed a 3D emotional space by adding either dominance
(degree of power over emotion) or liking/disliking. Although it is rare, a 4D
emotional space has also been studied by adding expectancy or naturalness.
While some researchers, e.g., \cite{Russell1980a},  argue that a 2D emotion
model is enough to characterize all categorical emotions, in this research, we
choose a 3D emotion model with valence, arousal, and dominance as the emotion
dimensions/attributes.

Darwin argued that the biological category of a species, like emotion
categories, does not have an essence due to the high variability of individuals
\citep{charles1872expression}. 
\cite{mehrabian1974approach} developed a pleasure,
arousal, and dominance (PAD) model to assess environmental perception,
experience, and psychological responses, as an alternative to categorical
emotion. The latter, also called as dimensional emotion, may represent human
emotion better than categorical emotion. This dimensional emotion view is also
known as the circumplex model of affect, and the pleasure dimension is often
replaced by valence for the same meaning (the VAD model).  Although most
research used the 2D model (valence and arousal), recent research shows four
dimensions needed to represent the meaning of emotion words
\citep{Fontaine2017}. However, current datasets lack the availability of
the fourth dimension label (i.e., expectancy). We evaluate the VAD emotion
model since the datasets also present the labels in 3D space.

Deep neural networks (DNN) have recently gained more interest in modeling human
cognitive processing for several tasks. \cite{Fayek2017} evaluated some DNN
architectures for categorical SER. They found fully connected (FC) networks and
recurrent neural networks (RNN) worked well for SER tasks using acoustic
features only. In neuropsychological science, the neural mechanism that
integrates acoustic (verbal) and linguistic (non-verbal) information remains
unclear \citep{Berckmoes2004}. The paper also stated that ``the various
parameters of prosody [acoustics] are processed separately in specific brain
areas" while no information is given for linguistic processing. In this
understanding, separation of acoustic and linguistic/text processing is better
modeled by a late fusion than an early fusion. This research makes use of a
support vector machine (SVM) for a late-fusion prediction from DNN-based
acoustic and linguistic emotion recognitions. The small remaining test data
after used by DNNs is a reason to use SVM over DNN.

This study aims to evaluate the combination of acoustic and text features to
improve the performance of dimensional automatic SER by using two-stage
processing. Current research on pattern recognition has also shown that the use
of multimodal features from audio, visual, and motion-capture data increases
performance as compared to using a single modality \citep{Hu2018, Yoon2018,
Tripathi2018}. Meanwhile, research on big data has revealed that the use of more
data will improve performance for results from the same algorithm
\citep{Halevy2009}. By using both acoustic and text features, SER should obtain
improved performance over acoustic-only and text-only recognition. This
assumption is also motivated by the fact that human emotion perception uses
multimodal sensing, peculiarly verbal and non-verbal information. Many
technologies, such as human-robot interaction, can potentially benefit from such
improvement in emotion recognition. 

The main contributions of this study then are: (1) a proposal of two-stage
processing for dimensional emotion recognition from acoustic and text features
using LSTM and SVM, and a comparison of the results with unimodal results and
another fusion method on the same metric and dataset scenario; (2) an
evaluation of different acoustic and text features to find the best pair of
acoustic-text pair based on evaluated features, including a frame-based
acoustic feature and utterance-based statistical functions with and without
silent pause features; (3) evaluation of speaker-dependent vs.
speaker-independent scenarios in dimensional speech emotion recognition from
text features; and (4) evaluation of using text features on a dataset that
originally contains target sentences but removed to avoid the effect of these
target sentences.

The rest of this paper is organized as follows. ``Related work" reviews closely
related work to this research, including the difference between this study and 
previous research, ``Datasets and features" outlines the datasets and feature
sets used in this research, ``Two-stage bimodal emotion recognition" explains
the method to achieve the results, ``Results and discussion" shows the results
and its discussion, and finally ``Conclusions" concludes this study and
proposes future work.

\section{Related work}
Speech emotion recognition (SER) began to be seriously researched as part of
human-computer interaction with the work by e.g., \cite{Kleine-cosack2006}. The
amount of research has grown as datasets have become publicly available,
including the Berlin EMO-DB, IEMOCAP, MSP-IMPROV, and RAVDESS datasets.  To
enable the analysis and comparison with previous research, we include the
following literature reviews of related work. We focus on comparing previous
work that used the same or similar datasets as this work does (specifically,
IEMOCAP, MSP-IMPROV, or both), and especially on research that focused on
dimensional rather than categorical emotion. While the focus here is on bimodal
emotion recognition using both acoustic and text data, some work on speech-only
or text-only emotion recognition is briefly described.

\subsection{Acoustic emotion recognition}
Recognition of emotion within speech signals has been actively developed since
the success of recognizing emotion via facial expressions. From categorical
emotion detection, the paradigm of SER has shifted to predicting degrees of
emotion attributes or dimensional emotions. One of the earliest papers on
(categorical) SER \citep{Petrushin1998} explored how well humans and computers
recognize emotion in speech. Since then, research on categorical emotion
recognition has grown following the development of affective research in
psychology.

\cite{Jin2005} reported a first trial on SER in categorical and two-dimensional
(2D) spaces. They found that acoustic features are helpful in describing and
distinguishing emotion through the concept of emotion modeling (2D space). In
2009, \cite{Giannakopoulos2009} re-investigated the association of speech
signals with an emotion wheel (continuous space). They proposed a method to
estimate the degrees of valence and arousal. Their method, including a proposed
feature set, could estimate both valence and arousal, with an error close to
that of average human annotation. \cite{Grimm2007a} used a fuzzy-logic
estimator and a rule base derived from acoustic features in speech, such as
pitch, energy, speaking rate, and spectral characteristics, to describe emotion
primitives (valence, arousal, and dominance). They obtained a moderate to high
correlation (0.42 \textless ~r \textless 0.85) between their method and human
annotation.

Using the IEMOCAP dataset, Parthasarathy and Busso tried to train a neural
network system to predict valence, arousal, and dominance simultaneously
\citep{parthasarathy2017jointly}. They proposed a multitask learning (MTL)
system based on the mean squared error (MSE) to balance the prediction of the
three emotion dimensions. They found that by combining a shared layer and an
independent layer, the MTL system's best performance exceeded that of the
traditional single-task learning (STL) method.

\cite{Abdelwahab2018} proposed using a domain-adversarial neural network (DANN)
to solve the problem of mismatch between training and test data in dimensional
SER. Using the DANN, they obtained performance that significantly improved 
that of a source-trained DNN. Thus, they addressed the importance of minimizing
the mismatch between the source (training) and target (test) data. Furthermore,
using the DANN showed that creating a flexible, discriminant feature
representation can reduce the gap in the feature space between the source and
target domains.

Some of the above results on dimensional SER showed that recognizing valence is
more difficult than recognizing arousal. To overcome this issue,
\cite{Sridhar2018} used higher regularization (dropout) for valence than for
the other dimensions when training SER through a DNN. Their system analysis
showed that a higher dropout is needed for predicting valence. By using higher
regularization, models could identify more general acoustic patterns that were
observed across speakers. \cite{Elbarougy2014} used a three-layer model based
on human perception for the same purpose. \cite{Li2019} improved on that work
by combining acoustic features for multilingual emotion recognition.

Although some improvements have been achieved, \cite{ElAyadi2011} addressed the
SER issue of whether it suffices to use acoustic features for modeling emotions
or it is necessary to combine them with other types of features, such as
linguistic discourse information or facial features. Text features can be
obtained through automatic speech recognition (ASR) and may be helpful in
significantly improving SER performance. In particular, text features are
expected to improve the performance of valence recognition, for which acoustic
features have typically failed to achieve high performance. Moreover, text
features are commonly used for sentiment analysis, which is similar to valence
prediction.

\subsection{Text emotion recognition}
As mentioned above, one area of research on text processing focuses on sentiment
analysis. This area is closely related to recognizing valence, i.e., the
polarity or semantic orientation of an event, object, or situation
\citep{Jurafsky2017}. Although early research sought to recognize sentiment
in text, extension to recognize categorical and dimensional emotion has been
attempted in recent years. As in other areas of research on pattern recognition,
some researchers in text processing have used unsupervised learning to detect
emotion in text \citep{Mantyla2016, Mohammad2016}, while others have used
supervised learning based on machine learning \citep{Alm2005, Yang2016}.

\cite{Alm2005} used a bag-of-words (BoW) model and other text features from text
datasets to predict emotion within those datasets. Using a multiclass linear
classifier, they obtained encouraging results that suggest a potential direction
for future research. Their proposed method could predict basic emotion from text
with an accuracy close to 70\%. 

\cite{Kim2010} used unsupervised learning to predict categorical emotion from
three different datasets: SemEval, ISEAR, and Fairy Tales. Using three
different techniques, they found that the best performance was achieved with
categorical classification based on non-negative matrix factorization (NMF).
\cite{Atmaja2019} used a deep-learning-based classification model and improved
the precision, recall, and F-score results for the ISEAR dataset from 0.528,
0.417, and 0.372 to 0.56, 0.54, and 0.54, respectively. They showed the
effectiveness of the deep-learning-based method for categorical emotion
recognition on a larger dataset, while on a smaller dataset, the unsupervised
approach achieved better results. Apart from categorical emotion recognition,
\cite{Atmaja2019} also performed dimensional emotion recognition on the same
dataset used for the categorical task. Similar to the categorical task, the
results showed fewer errors when the size of the training set was increased.

\cite{Mantyla2016} used emotion words from an affective lexicon to mine valence,
arousal, and dominance in text communication. Specifically, they used text
communication data from a software development situation, including issues and
comments captured through issue repository technology. They used the measure of
valence, arousal, and dominance (VAD) to detect the productivity and burnout of
the software developers. The results showed that increased emotions in terms of
VAD correlated with increased productivity. Their results also complemented
previous results showing that VAD can be measured from text, though at first,
only the sentiment (valence) was used in text processing.

Research on text emotion recognition has usually used written language (from
chats, Twitter, forum threads, etc.), which differs from spoken language. Also,
most such work on text processing has detected only the valence, i.e., only one
emotion dimension. Because speech transcription converts spoken language to a
written form, it should contain more emotional information than written plain
text. Evaluation of the other emotion dimensions (arousal and dominance) is
also necessary to determine the impact on those dimensions, along with
evaluation of the combination with acoustic features for that purpose.

\subsection{Bimodal emotion recognition}
Using bimodal or multimodal features for emotion recognition is not new. Among
many modalities, audio and visual features are the most used for extracting
emotion information. When only speech is conveyed, however, two types of
information can be extracted: acoustic and text features. Among many research
papers, the reports by \cite{Eyben2010}, \cite{Karadogan2012}, \cite{Ye2014},
\cite{Jin2015}, \cite{Aldeneh2017}, \cite{Yoon2018}, \cite{Atmaja2019b}, and
\cite{Zhang2019} are the most related to this paper.

\cite{Eyben2010} proposed an online method to detect not only valence and
arousal but also the time when those emotion attributes are detected. They used
a recurrent neural network (RNN) based on long short-term memory (LSTM) to
recognize a framewise valence-arousal continuum with time. By adding a keyword
spotter, they were able to improve the performance by using regression
analysis. The results were measured in Pearson correlation coefficient (PCC).
They also found that keywords like ``again,'' ``angry,'' ``assertive,'' and
``very'' were related to activation, while typical keywords correlated to
valence were ``good,'' ``great,'' ``lovely,'' and ``totally.'' Similar to
that idea, \cite{Karadogan2012} used affective words from Affective Norms for
English Words (ANEW) to determine a valence-arousal value and combine it with
a result from acoustic features. The latter paper also obtained similar
improvement over using a single modality.

\cite{Ye2014} used bimodal features from acoustic and text information to
recognize emotion within speech. The acoustic features were trained in two
parallel classifiers: an SVM and a backpropagation network. The text features
were trained in two serial classifiers, which were both Naive Bayes classifiers.
The second classifier acted as a filter for unreliable parts from the first
classifier. Decision-level fusion (late fusion) was then implemented by
combining the acoustic and text features with tree-weighting factors for the
SVM, backpropagation network, and text classifiers. The resulting fusion method
obtained 93\% accuracy, as compared to 83\% from the acoustic features only and
89\% from the text features only. The task was categorical emotion detection
from a Chinese database. Similar to that approach for a categorical task,
\cite{Jin2015} used the IEMOCAP dataset to test combinations of acoustic and
text features for SER. The novelty of their method was the use of an emotion
vector for lexical features, which improved the accuracy in four-class emotion
recognition from 53.5\% (acoustic) and 57.4\% (text) to 69.2\% (acoustic +
text).

\cite{Aldeneh2017} used acoustic and lexical features to detect the degree of
valence from speech. They used 40 mel-filterbanks (MFBs) as acoustic features
and word vectors as text features. Continuous valence values were then converted
to three categorical classes: negative, neutral, and positive. Using that
approach, they improved the weighted accuracy from 64.5\% (text) and 58.9\%
(acoustic) to 69.2\% (acoustic + text). 

\cite{Yoon2018} used audio and text networks to predict emotion classes from the
IEMOCAP dataset. Both networks used RNNs with inputs of mel-frequency cepstral
coefficients (MFCCs) for audio and word vectors for text. The proposed
multimodal dual recurrent encoder (MDRE) improved on the single-modality RNNs
from 54.6\% (audio) and 63.5\% (text) to 71.8\% (audio + text).
\cite{Atmaja2019b} obtained a better result by using 34 acoustic features after
silence removal and combining them with word embeddings. With  LSTM used for the
text and dense networks for speech, the latter paper obtained an accuracy of
75.49\% on the same dataset and task.

Instead of using lexical features, \cite{Zhang2019} used phonemes and combined
them with acoustic features to recognize valence in speech. They used 39 unique
phonemes from the IEMOCAP and MSP-IMPROV datasets and a
40-dimensional log-scale MFB energy for the acoustic features. Using a scaled
version of valence, converted from a 5-point scale to three categorical
classes, they showed that their multistage fusion model outperformed all other
models on both IEMOCAP and MSP-IMPROV.

\subsection{Multimodal emotion recognition}
The term multimodal used in this subsection refers to the use of three or
more modalities for emotion recognition. These modalities are usually visual,
audio, text, gesture, and eye gaze. In a recent survey \citep{Poria2017}, it is
confirmed that the use of multimodal classifiers can outperform unimodal
classifiers. Motivated by human multimodal experiences, multimodal processing
not only applies to emotion recognition but also for other applications,
e.g., audio-visual speech recognition, event detection, and media
summarization.

\cite{Zadeh2017} used audio, visual, and text for sentiment analysis. They
proposed a model, termed a tensor fusion network (TFN), by posing
intra-modality and inter-modality dynamics of multimodalities. The proposed
method represents interactions between unimodal, bimodal, and trimodal
features. The experiment on the publicly available CMU-MOSI dataset produced
state-of-the-art results for sentiment analysis.

\cite{Tripathi2018} used audio, text, and motion capture (mocap) to identify
the categorical emotion of the IEMOCAP dataset. Using unimodal features, they
obtained an accuracy of 55.65\%, 64.78\%, and 51.11\% for audio, text, and
mocap, respectively. Using multimodal features, they improved the accuracy of
71.04\%. They concatenated different classifiers (LSTM for text and audio, CNN
for mocap) to obtain the final categorical emotion prediction.

The lack of multimodal emotion recognition itself is the necessity 
of several modalities, mainly audio and video. In some cases, only audio
data are available, e.g., telephone calls, voice assistants, and audio
messages. Due to privacy and other reasons, the use of videos and other
modalities may limit obtained data for predicting emotional states. In the case
of speech, acoustic and linguistic features can be extracted back to obtain
bimodal features.

Apart from the advantages and disadvantages of bimodal/multimodal recognition
over using a single modality, there is a need to develop a new method for SER.
The reasons are (1) some prior research did not predict all emotional attributes
\citep{Eyben2010, Karadogan2012, Zhang2019}, while other studies predicted
emotion categories instead of attributes; (2) instead of predicting continuous
emotion attribute scores, some studies switched to a categorical task for
simplicity \citep{Zhang2019, Aldeneh2017}; and (3) some of the reported results
are not up to date and showed low improvement \citep{Eyben2010, Karadogan2012}.

Although we have limited our work to using both acoustic and text features,
other researchers have already proposed another solution to solve the issues
above, namely, using audiovisual emotion recognition. Nevertheless, there is
still a need to propose and evaluate methods using both acoustic and text
features because some target applications only involve speech data. In these
voice-based applications, no visual or written-text information is acquired. To
maximize the resources for extracting emotion within speech, this paper exploits
both acoustic and text features (obtained via speech transcription) and combines
them for dimensional emotion regression. By using both kinds of information in a
two-stage process, we expect the proposed method's performance to be close to or
exceed the performance obtained by using visual information. Note here that
visual information, particularly facial expressions, has been reported to have
more influence on dimensional emotion than other modalities do
\citep{FabienRingeval2018}.

\section{Data and feature sets}
\subsection{Datasets}
Datasets for investigating our proposal to use two-stage processing for
dimensional SER must meet certain requirements. The requirements are that (1)
the dataset has both speech data and text transcription (to speed up text data
acquisition), (2) the dataset is already annotated with dimensional labels, and
(3) the dataset is publicly available. The following two datasets satisfy these
requirements.

\noindent 1. IEMOCAP  \\
IEMOCAP, which stands for interactive emotional dyadic motion capture database,
contains recordings of dyadic conversations with markers on the face, head, and
hands. The recordings thus provide detailed information about the actors's
facial expressions and hand movements during both scripted and spontaneous
spoken communication scenarios \citep{Busso2008}. This research only uses the
acoustic and text features because the goal is bimodal speech emotion
recognition. The IEMOCAP dataset is freely available upon request, including
its labels for categorical and dimensional emotion. We use the dimensional
emotion labels, which are average scores for two evaluators because they
enable deeper analysis of emotional states. The dimensional emotion scores, for
valence, arousal, and dominance, are meant to range from 1 to 5 as a result of
Self-Assessment Manikin (SAM) evaluation. We have found some labels with scores
lower than 1 or higher than 5; however, we remove those data (seven samples).
All labels are then converted from the 5-point scale to a floating-point values
in range [-1, 1] when they are fed to a DNN system.

The total length of the IEMOCAP dataset is about 12 hours, or 10039
turns/utterances, from ten actors in five dyadic sessions (two actors each). The
speech modality used to extract acoustic features is a set of files in the
dataset with a single channel per sentence. The sampling rate of the speech data
was 16 kHz. For text data, we use manual transcription in the dataset
without additional preprocessing.

\noindent 2. MSP-IMPROV \\
MSP-IMPROV, developed by the Multimodal Signal Processing (MSP) Lab at the
University of Texas, Dallas, is a multimodal emotional database obtained by
applying lexical and emotion control in the recording process while also
promoting naturalness. The dataset provides audio and visual recordings, while
text transcriptions are obtained via automatic speech recognition (ASR) provided
by the authors. As with IEMOCAP, we use speech and text data with
dimensional emotion labels. The annotation method for the recordings was the
same as for IEMOCAP, i.e., SAM evaluation, with rating by at least five
evaluators. We treat missing evaluations as neutral speech (i.e., a score of 3
for valence, arousal, and dominance). Also as with IEMOCAP, all labels are
converted to floating-point values in a range [-1, 1] from the original 5-point
scale.

The MSP-IMPROV dataset was designed within a dialogue framework to elicit
target sentences that had the same semantic content but were produced with
different emotional expressions. In one recording, the target sentences were
produced ad lib; for another recording, the target sentences were read.  These
two recordings are referred to as ``Target-improvised" and ``Target-read",
respectively. For our purposes, since our goal is to examine the effect of both
linguistic and acoustic information on emotional ratings, these recordings were
not appropriate for our study. However, we were able to use two sets of
recordings that did not have the same semantic content, which is called
``Other-improvised" and ``Natural-interaction". The former included
conversations of the actors during improvisation sessions;
the latter included the exchanges during the breaks, while the actors
were not acting, which also were being recorded. A similar protocol was used by
\cite{Zhang2019}, and we followed their lead in referring to this subset of the
MSP-IMPROV dataset as MSP-I+N (MSP improvised and natural interaction), or
MSPIN. In our work, we included the text transcriptions used by Zhang et al.
(transcriptions are provided by the authors of the dataset);
for the additional utterances not included in the Zang study, transcriptions
were obtained using Mozilla's DeepSpeech \citep{DeepSpeech2019}.  We thus use
7166 utterances from a total of 8438. The speech data in the dataset was
sampled in mono at 44.1 kHz, with one file per utterance/sentence.

\begin{figure}
\includegraphics[width=5in]{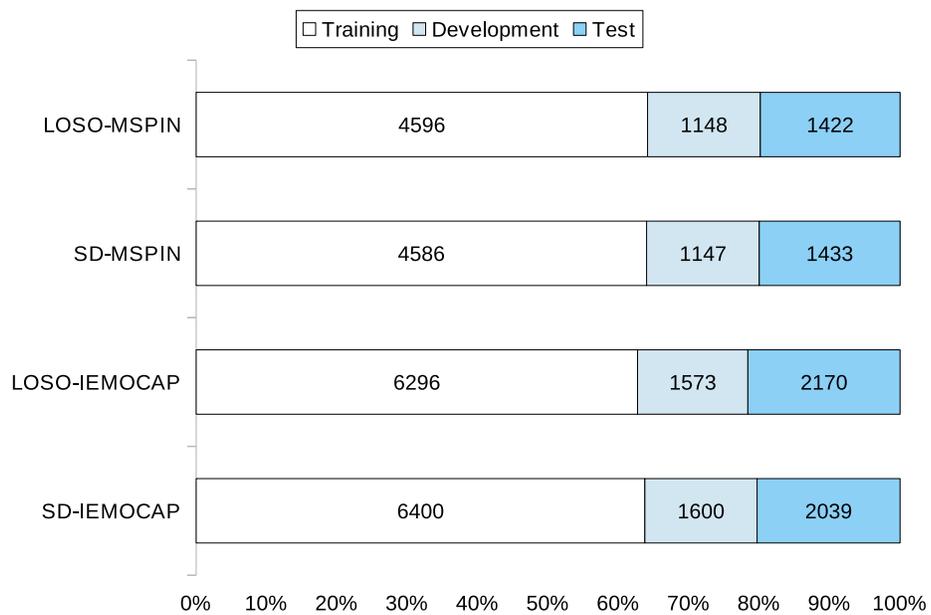}
\caption{Proportions of data splitting for each partition of each dataset. In
one-stage LSTM processing, the outputs of the model are both development and
test data. In the second stage SVM processing, the input data is the prediction
from the development set of the previous stage and the output is the prediction
of test data.}
\label{fig:data_partition}
\end{figure}

We split each dataset into two partitions to observe any differences between a
speaker-dependent (SD) partition and a speaker-independent partition made by
leaving one session out (LOSO) for each dataset. For example, for the IEMOCAP
dataset, the last session (i.e., session 5), which is recorded from two
different actors (out of 10) is only used for testing. Similarly, for MSP-I+N,
all utterances from session 6 (two speakers out of 12) are used for the test
set.  Our rule for data splitting is to divide between the training +
development and test sets in a ratio close to 80:20. This rule is applied for
both the SD and LOSO partitions. Then, of the training + development data, 80\%
is used for training and the remaining 20\% is used for development, as shown
in Figure \ref{fig:data_partition}. Both methods are evaluated with the same
unseen test sets to compare the performance and measure the improvement. Note
that we did not use cross validation (but instead divided into training and
test data) for evaluation since the number of samples for both datasets is
adequate (10039 and 7166 samples).  This strategy is also utilized to keep the
same test set for LSTM (one-stage processing) and SVM (two-stage processing)
which is difficult if the samples are shuffled/cross-validated.

\subsection{Feature sets} 
\label{sect:feature_sets}
Most research on SER, or generally on pattern classification, focuses on two
main topics: feature extraction, as in \cite{Batliner2011}, and
classification/regression methods, as in \cite{Albornoz2011}. While this
research focuses on the second topic, we also evaluate state-of-the-art feature
sets used for SER. For acoustic features, we evaluate three feature sets: the
Geneva Minimalistic Acoustic Parameter Set (GeMAPS), statistical functions from
GeMAPS, and the same functions from GeMAPS with a silence feature. For text
features, aside from the original word vectors extracted from the text
transcription, we also evaluate two word embeddings that are pretrained on a
larger corpus: the Word2Vec embedding \citep{Mikolov} and GloVe embedding
\citep{Pennington2014}. These feature sets are explained below.

\paragraph{Acoustic features} The type of acoustic features extracted from a
speech signal is the most important part of an SER system. GeMAPS is an effort
to standardize the acoustic features used for voice research and affective
computing \citep{Eyben}. The feature set consists of 23 acoustic low-level
descriptors (LLDs) such as fundamental frequency ($f_0$), jitter, shimmer, and
formants, as listed in Table \ref{tab:aco_feature}. As an extension of GeMAPS,
eGeMAPS includes statistical functions derived from the LLDs, such as the
minimum, maximum, mean, and other values. Since these features are extracted on
frame-based processing, the feature size becomes large for one utterance
(e.g., 3409 $\times$ 23 for IEMOCAP), which is suitable for deep learning
methods like LSTM. Including the LLDs, the total number of features in eGeMAPS
is 88. These statistical values are often called high-level statistical
functions (HSF). \cite{Schmitt2018} found, however, that using only the mean and
standard deviation (std) from the LLDs achieved a better result than using
eGeMAPS and audiovisual features. These global features may represent more
emotion information within speech than frame-based features. We thus coded these
two statistical functions (47 values) from the LLDs as the HSF1 feature set. We
also investigate the effect of including a silence feature in this SER research,
as explained below. We define the combination of HSF1 with the silence feature
as HSF2.

Silence, in this paper, is defined as the proportion of silent frames among all
frames in an utterance. In human communication, the proportion of silence in
speaking depends on the speaker's emotion. For example, a happy speaker may have
fewer silences (or pauses) than a sad speaker. The proportion of silence in an
utterance can be calculated as

\begin{equation}
    \label{eq:sil_ratio}
 S = \frac{N_{s}}{N_{t}},
\end{equation}
where $N_s$ is the number of frames categorized as silence (silent frames), and
$N_t$ is the total number of frames. A frame is categorized as silent if it does
not exceed a threshold value defined by multiplying a factor by a root mean
square (RMS) energy, $X_{rms}$. Mathematically, this is formulated as

\begin{equation}
 th = \alpha \times \overline{X_{rms}},
\end{equation}
where $X_{rms}$ is defined as

\begin{equation}
 X_{rms} = \sqrt{\frac{1}{n}\sum_{i=1}^{n}x[i]^2}.
\end{equation}

\begin{figure}[htpb]
\centering
\includegraphics[width=0.8\textwidth]{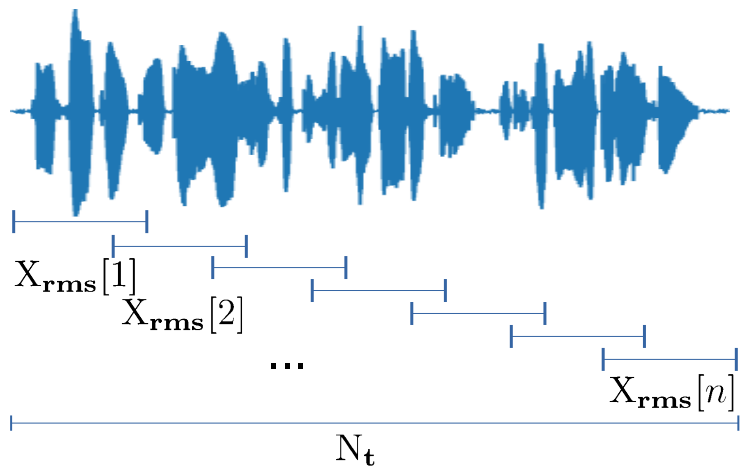}
\caption{Moving frame to calculate the silence feature.}
\label{fig:silence_fig}
\end{figure}

This silence feature is similar to the disfluency feature proposed in
\cite{moore2014word}. In that paper, the author divided the total duration of
disfluency by the total utterance length for $n$ words. Figure
\ref{fig:silence_fig} illustrates the calculation of our silence feature. If
$X_{rms}$ from a frame is below $th$, then it is categorized as silent, and the
calculation of equation \ref{eq:sil_ratio} is applied.

\begin{table}
 \centering
 \caption{Acoustic feature sets derived from the GeMAPS features by \cite{Eyben} and the statistical functions used for dimensional SER in this research.}
 \begin{tabular}{p{5cm} p{2.9cm} p{2.9cm}}
 \hline
 LLDs & HSF1 & HSF2 \\
 \hline
intensity, alpha ratio, Hammarberg index, spectral slope 0-500 Hz, spectral
slope 500-1500 Hz, spectral flux, 4 MFCCs, F0, jitter, shimmer,
harmonics-to-noise ratio (HNR), harmonic difference H1-H2, harmonic difference
H1-A3, F1, F1 bandwidth, F1 amplitude, F2, F2 amplitude, F3, and F3 amplitude.
& mean (of LLDs), standard deviation (of LLDs) & mean (of LLDs), standard
deviation (of LLDs), silence \\
 \hline
 \end{tabular}
 \label{tab:aco_feature}
\end{table}

\paragraph{Text features} To process a word sequence in a computational model,
the text must be converted to numerical values. The resulting text feature is
commonly known as word embedding and is a vector representation of a word.
Numerical values in the form of a vector are used to enable a computer to
process text data, as it can only process numerical values. The values are
points (numeric data) in a space whose number of dimensions is equal to the
vocabulary size. The word representations embed those points in a feature space
of lower dimension. In the original space, every word is represented by a
one-hot vector, with a value of 1 for the corresponding word and 0 for other
words. The element with a value of 1 is converted to a point in the range of the
vocabulary size. 

In addition to directly converting the text in the transcriptions of the
datasets (IEMOCAP and MSP-I+N) to sequences, two pretrained word embeddings are
used to weight the original word embeddings. As mentioned above, the two
word-embedding models are Word2Vec \citep{Mikolov} and GloVe
\citep{Pennington2014}. Hence, we have three different text features for word
embeddings: (1) a text sequence, or word embedding (WE), without any weighting,
(2) a WE weighted by a pretrained Word2Vec model, and (3) a WE weighted by a
pretrained GloVe embedding model. Word2Vec is trained on large datasets to
model its vector representation by using either a continuous bag-of-words or a
continuous skip-gram model. GloVe is another model to represent vector in words
by analyzing linear direction by using a global log-bilinear regression method.
All these features are fed into the same embedding layer in the text network.

\section{Two-stage bimodal emotion recognition}
\subsection{Acoustic emotion recognition system}
Most SER research uses only acoustic features. Our approach to acoustic SER is
similar to that research. The contribution of our acoustic network is the
evaluation of mean + std features at the utterance level and the use of a
silence feature with the statistical functions to investigate any improvement.
This evaluation is a continuation of \citep{Atmaja2020f} with extension 
on different feature sets and datasets.
As explained in section \ref{sect:feature_sets}, we evaluate three acoustic
feature sets: LLDs, HSF1, and HSF2.

The LLD features are the 23 acoustic features listed in Table
\ref{tab:aco_feature}. For each frame (25 ms), these 23 acoustic features are
extracted. With a hop size of 10 ms, the maximum number of sequences is 3409 for
the IEMOCAP dataset and 3812 for the MSP-I+N dataset. Hence, the size of the
input is 3409 $\times$ 23 for IEMOCAP and 3812 $\times$ 23 for MSP-I+N. The
extraction process uses the openSMILE toolkit \citep{Eyben2016open}.

\begin{figure}[htpb]
\centering
\includegraphics[width=0.8\textwidth]{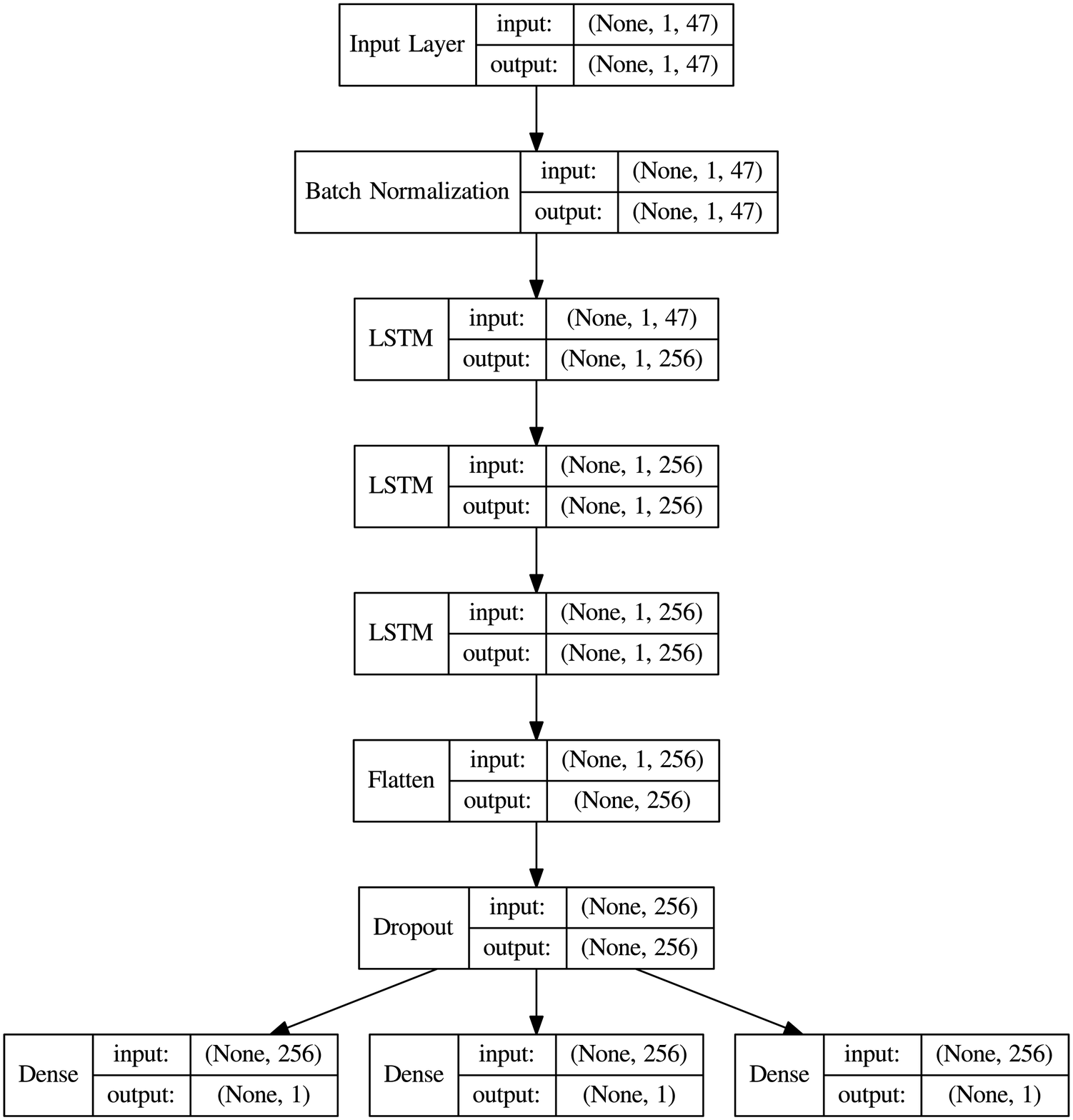}
\caption{Structure of acoustic network to process acoustic features.}
\label{fig:acoustic_model}
\end{figure}

Figure \ref{fig:acoustic_model} shows an overview of the acoustic network.
LSTM is chosen because the number of training samples is adequate ($> 5000$
samples) and it shows good results in the previous research
\citep{Schmitt2018}. Before entering the LSTM layers, the LLD features at the
input layer are fed into a batch normalization layer to speed up the
computation process. The three subsequent LSTM layers are stacked with 256
nodes in each layer, following one of the configurations in 
\cite{Abdelwahab2018}. Instead of returning the last
output of the last LSTM layer, we designed the network to return the full
sequence and flatten it before inputting it into three dense layers that
represent valence, arousal, and dominance. The outputs of these last dense
layers are then the predictions for those emotion attributes, i.e., the degrees
of valence, arousal, and dominance in the range [-1, 1]. 

The tuning of hyper-parameters follows the previous research
\citep{Atmaja2019b,Atmaja2020d}. A batch size of 8 was used with a maximum of
50 epochs. An early stop criterion with ten patiences stops the training
process if no improvement were made in 10 epochs (before the maximum epoch).
The last highest-score model was used to predict the development data. An
RMSprop optimizer was used with its default learning rate, i.e., $0.001$. Table
\ref{tab:dnn_params} shows the setups on acoustic and text networks. These
setups were obtained based on experiments with regard to the size of networks.
For instance, the smaller acoustic networks with HSF features employed $\tanh$
output activation function and did not use the dropout rate, while the larger
acoustic networks (with LLD) and text networks employed linear activation
function and dropout rate.

\begin{table}
    \caption{The hyper-parameter used in experiments}
    \begin{center}
        \begin{tabular}{l| c c}
            \hline
            Hyper-parameter  &  Acoustic network    & Text network \\
            \hline
            network type        &   LSTM            &   LSTM \\
            number of layers    &   3               &   3 \\
            number of units     &   256             &   256 \\
            fourth layer        &   Flatten         &   Dense \\
            hidden activation   &   linear          &   linear \\
            output activation   & linear (LLD) / tanh (HSF)   &   linear \\
            dropout rate        & 0.3 (LLD) / 0 (HSF)&   0.3 \\
            learning rate       &   0.001           &   0.001  \\
            batch size          &   8               &   8  \\
            maximum epochs      &   50              &   50 \\
            optimizer           &   RMSprop         &   RMSprop \\
            \hline
        \end{tabular}
    \end{center}
    \label{tab:dnn_params}
\end{table}

For the HSF1 and HSF2 inputs on acoustic networks, the same setup applies.
These two feature sets are very small as compared to the LLDs: HSF1 has a size
of 1 $\times$ 46, while HSF2 has a size of 1 $\times$ 47. This big difference
in input size (1:1800) leads to faster computation on HSF1 and HSF2 than on the
LLDs. Note that although Figure \ref{fig:acoustic_model} shows HSF2 as the
input feature, the same architecture also applies to the LLDs and HSF1.

The idea of using LSTM is to hold the last output in memory and use that output
as a successive step. For instance, LLD with ($3409, 23$) feature size will
process the first time step 1 to the last time step 3409. For HSF1 and HSF2,
which contains a single time stamp, the data is processed only once ([1, 46]
and [1, 47] for HSF1 and HSF2). Here, the only difference from multi-time
steps is that the network performs three passes (forget gate, input gate, and
output gate) instead of a single pass (see \cite{Eyben2010}).  This information
will include all information from the networks' memory.

\subsection{Text emotion recognition system}
\begin{figure}[htpb]
\centering
\includegraphics[width=0.8\textwidth]{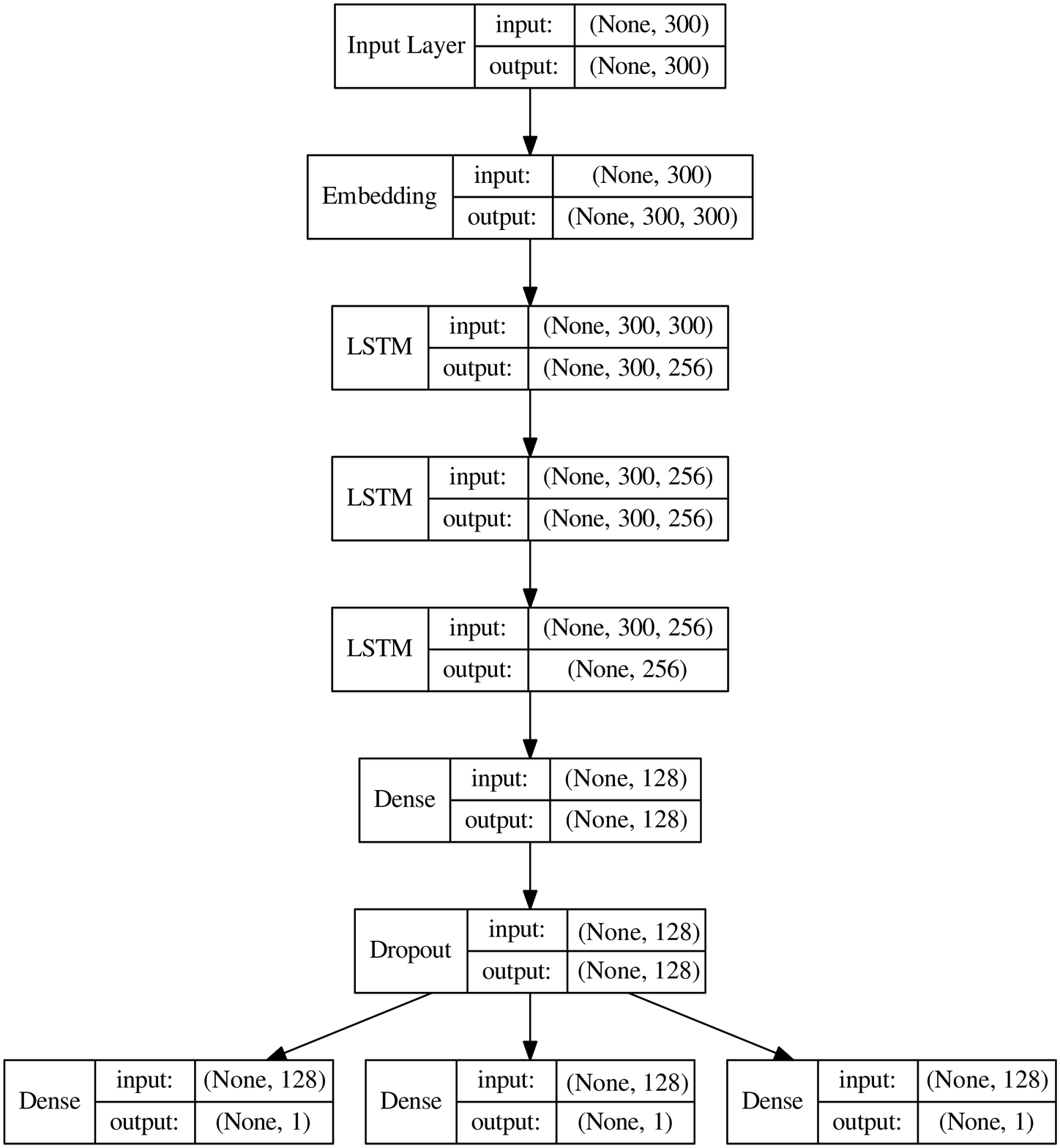}
\caption{Structure of text network to process word embeddings/vectors.}
\label{fig:text_model}
\end{figure}

The text network, shown in Figure \ref{fig:text_model} for the MSP-I+N dataset,
uses the same input size for the three different text features. The WE, WE with
pretrained Word2Vec, and WE with pretrained GloVe embedding have 300
dimensions for each word. The longest sequence in the IEMOCAP dataset is 100
sequences (words), while for MSP-I+N, the longest is 300 sequences. Hence, the
input feature sizes for the LSTM layers are 100 $\times$ 300 for IEMOCAP and
300 $\times$ 300 for MSP-I+N with its corresponding number of samples. The same
three LSTM layers are stacked as in the acoustic network, but the last LSTM
layer only returns the last output. A dense layer with a size of 128 nodes is
added after the LSTM layers and before the last three dense layers. Between the
dense layers is a dropout layer with the same probability of 0.3 to avoid
overfitting.

\subsection{Multitask learning}
\label{sub:mtl}
The task here in dimensional emotion recognition is to simultaneously predict
the degrees of three emotion attributes, i.e., the degrees of valence, arousal,
and dominance, for any given utterance. As the main target metric is the
concordance correlation coefficient (CCC), the loss function is the CCC loss
(CCCL). CCC also measures the performance of regression analyses by comparing
gold standard labels and predictions. CCCL computes the score difference
between the labels and predicted values for the three attributes. The CCC and
CCCL are formulated as the following:
\begin{align} 
CCC &= \dfrac{2 \rho \sigma_x \sigma_y} {\sigma_x^2 + \sigma_y^2 + (\mu_x - \mu_y)^2}, \\
CCCL &= 1 - CCC,
\end{align}
where $\rho$ is the Pearson correlation coefficient between the predicted
emotion degree $x$ and the true emotion degree $y$, $\sigma^2$ is the variance,
and $\mu$ is the mean. As the learning process minimizes three variables, we use
the following multitask learning approach to optimize the CCC score:
\begin{equation}
CCCL_{tot} = \alpha CCCL_{V} + \beta CCCL_{A} + (1-\alpha-\beta) CCCL_{D},
\end{equation}
where $\alpha$ and $\beta$ are respective CCCL parameters for valence (V) and
arousal (A). The parameter for dominance (D) is obtained by subtracting $\alpha$
and $\beta$ from 1. The same parameter range [0, 1] with 0.1 steps is
investigated for $\alpha$ and $\beta$ for both the acoustic and text networks,
resulting in different optimal parameters, which are obtained by using linear
search. Note that only positive values of $CCCL_{D}$'s parameters are used to
investigate the optimal parameters.

\subsection{SVM-based late fusion}
\begin{figure}
\includegraphics[width=\textwidth]{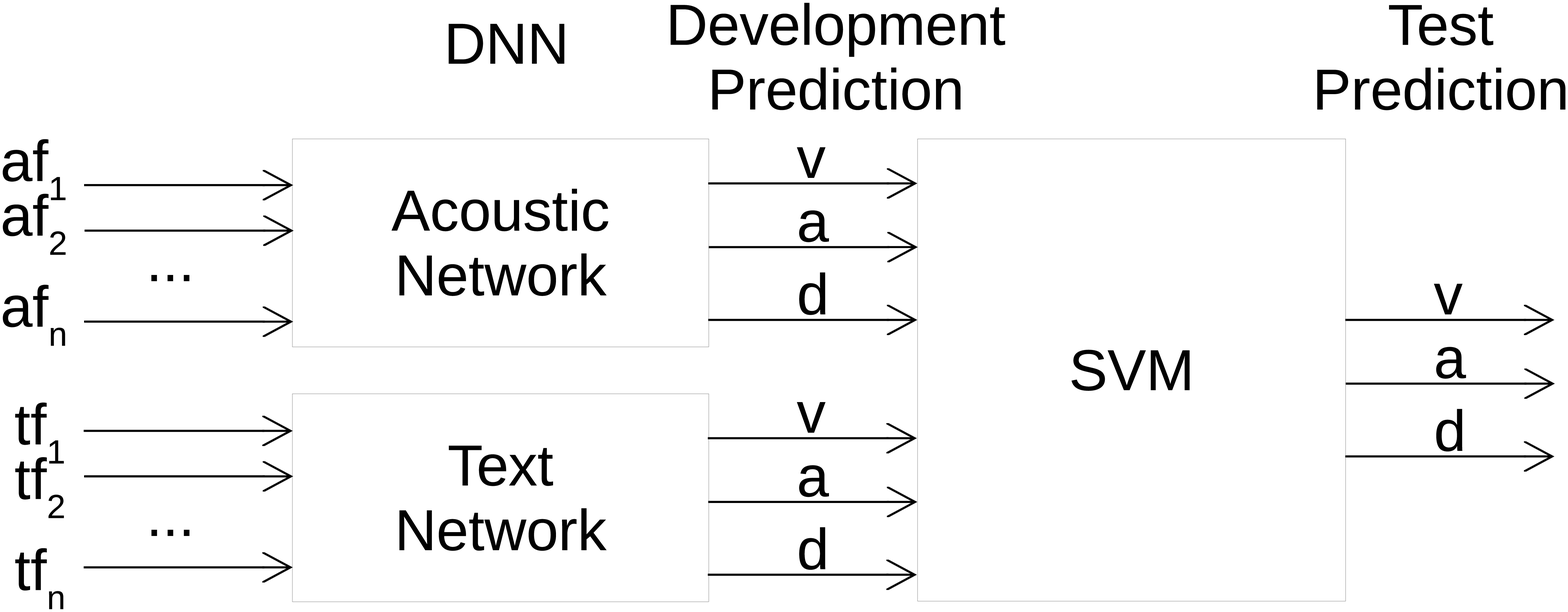}
\caption{Proposed two-stage dimensional emotion recognition method using DNNs 
and an SVM. The inputs are acoustic features (af) and text features (tf); 
the outputs are valence (v), arousal (a), and dominance (d).}
\label{fig:csl_system}
\end{figure}

We choose an SVM as the final classifier to fuse the outputs of the acoustic
and text networks because of its effectiveness in handling smaller data (as
compared to a DNN) and its computation speed. The datapoints produced by LSTM
processing as the input of SVM are small; i.e., 1600, 1538, 1147, and 1148 for
IEMOCAP-SD, IEMOCAP-LOSO, MSPIN-SD and MSPIN-LOS0, respectively. The SVM then
applies a regression analysis to map them to the given labels. Figure
\ref{fig:csl_system} shows the architecture of this two-stage emotion
recognition system using DNNs and an SVM. Each prediction from the acoustic and
text networks is fed into the SVM.  From two values (e.g., valence predictions
from the acoustic and text networks), the SVM learns to generate a final
predicted degree (e.g., for valence). The concept of using the SVM as the final
classifier can be summarized as follows.

Suppose that two valence prediction outputs from the acoustic and text networks,
$x_i = [x_{ser}[i], x_{ter}[i]]$, are generated by the DNNs, and that $y_i$ is
the corresponding valence label. The problem in dimensional SER fusing acoustic
and text results is to minimize the following: 

\begin{equation}
\begin{aligned}
& \underset{w, b, \zeta, \zeta^*}{\text{min}}
& & \frac{1}{2} w^Tw + C \sum_{i=1}^n \zeta_i + C \sum_{i=1}^n \zeta_i^* \\
& \text{subject to}
& & w^T \phi (x_i)+b- y_i \leq \epsilon  + \zeta_i, \\
&&& y_i - w^T \phi (x_i) -b \leq \epsilon + \zeta_i^*, \\ 
&&& \zeta_i, \zeta_i^* \geq 0, i = 1, \ldots, n,
\end{aligned}
\end{equation}
where $w$ is a weighting vector, $C$ is a penalty parameter, $\zeta$ and
$\zeta^*$ is the distance between misclassified points and the corresponding
marginal boundary (above or below).  Here, $\phi$ is the kernel function. We
choose a radial basis function (RBF) kernel because of its flexibility in
modeling a nonlinear process with a dimensional emotion model close to this
kernel. The function $\phi$ for the RBF kernel is formulated as

\begin{equation}
 K(x_i, x_j) = e^{\gamma(x_i - x_j)^2},
 \label{tab:label}
\end{equation}
where $\gamma$ defines how much influence a single training has on the model.
All parameters in this SVM are obtained empirically via linear search in a
specific range. Although the explanation above uses valence, the same also
applies for arousal and dominance. 

\subsection{Reproducibility}
The experimental code was written in Python, and, for the sake of research
reproducibility, it is available in the following repository:
\url{https://github.com/bagustris/two-stage-ser}. The DNN part was implemented
using Keras by \cite{chollet2015keras} and Tensorflow, while the SVM-based
fusion was implemented using the scikit-learn toolkit by \cite{scikit-learn}. To
obtain consistent results for each run, some fixed numbers are initialized at
the beginning, as can be found in the repository above.

\section{Results and discussions}
\subsection{Results from single modality}
Before presenting the bimodal feature-fusion results, it is important to show
the results of unimodal emotion recognition. The goals here are (1) to observe
the (relative) improvement of bimodal feature fusion over using a single
modality, and (2) to observe the effects of different features on different
emotion attributes.

Tables \ref{tab:ser-test} and \ref{tab:ter-test} list the single-modality
results of dimensional emotion recognition from the acoustic and text networks,
respectively. In general, acoustic-based SER gave better results than text-based
SER in terms of the average CCC score. For particular emotion attributes, the
text network gave a higher CCC score for valence prediction than those obtained
by the acoustic network, except on the MSPIN datasets. This confirms the
previous finding by \cite{Karadogan2012} that valence is better estimated by
semantic features, while arousal is better predicted by acoustic features. In
addition, we found that the dominance dimension was better predicted by acoustic
features than by text features. This finding can be inferred from both tables,
in which the CCC scores for the dominance dimension are frequently higher from
the acoustic network than from the text network.

The exception of a higher valence score on the MSPIN-SD dataset by the acoustic
network can be seen as the effect of either the DNN architecture or the
dataset's characteristics. In \cite{chen2017multimodal}, the obtained score was
higher for valence than for arousal or liking (the third dimension, instead of
dominance) with their strategy on acoustic features. In contrast,
\cite{Abdelwahab2018} obtained a lower score for valence than for arousal and
dominance by using their proposed DANN method on the same MSP-IMPROV dataset
(whole data, all four scenarios). Given this comparison, we conclude that the
higher valence score obtained here was an effect of the DNN architecture
because of the multitask learning. Our result on a single modality (acoustic
network) outperformed the DANN result on MSP-IMPROV, where their highest CCC
scores were (0.303, 0.176, 0.476) as compared to our scores of (0.404, 0.605,
0.517) for valence, arousal, and dominance, respectively.

\begin{table}[htpb]
 \centering
 \caption{CCC score results of dimensional emotion recognition using an acoustic
     network. The best results on the test set are in bold. LLDs: low-level
     descriptors from GeMAPS \citep{Eyben}; HSF1: mean + std of LLDs; HSF2: mean
     + std + silence.}
 \label{tab:ser-test}
 \begin{tabular}{l c c c c}
 \hline
 Feature set & V & A & D & Mean \\
 \hline
 \multicolumn{5}{c}{IEMOCAP-SD} \\
 LLD	& 0.153	& 0.522	& 0.534	& \textbf{0.403} \\ 
 HSF1	& 0.186	& 0.535	& 0.466	& 0.396 \\
 HSF2	& 0.192	& 0.539	& 0.469	& 0.400 \\
 \hline
 \multicolumn{5}{c}{MSPIN-SD} \\ 
 LLD	& 0.299	& 0.545	& 0.441	& 0.428 \\
HSF1	& 0.400	& 0.603	& 0.506	& 0.503 \\
HSF2	& 0.404	& 0.605	& 0.517	& \textbf{0.508} \\
 \hline
 \multicolumn{5}{c}{IEMOCAP-LOSO} \\
 LLD	& 0.168	& 0.486	& 0.442	& 0.365 \\
 HSF1	& 0.206	& 0.526	& 0.442	& 0.391 \\
 HSF2	& 0.204	& 0.543	& 0.442	& \textbf{0.396} \\ 
 \hline
 \multicolumn{5}{c}{MSPIN-LOSO} \\
LLD	    & 0.176	& 0.454	& 0.369	& 0.333 \\ 
HSF1	& 0.201	& 0.506	& 0.357	& \textbf{0.355} \\
HSF2	& 0.206	& 0.503	& 0.346	& 0.352 \\
 \hline
 \end{tabular}
\end{table} 

\begin{table}[htpb]
 \centering
 \caption{CCC score results of dimensional emotion recognition using text
     networks; each score is an averaged score of 20 runs with its standard
     deviation. WE: word embedding; Word2Vec: WE weighted by pretrained word
     vector \citep{Mikolov}; GloVe: WE weighted by pretrained global vector
     \citep{Pennington2014}.}
 \label{tab:ter-test}
 \begin{tabular}{l c c c c}
 \hline
 Feature set & V & A & D & Mean \\
 \hline
 \multicolumn{5}{c}{IEMOCAP-SD} \\
WE	        & 0.389 $\pm$ 0.008 & 0.373 $\pm$ 0.010 & 0.398 $\pm$ 0.017 &	0.387 $\pm$ 0.010 \\
Word2Vec	& 0.393 $\pm$ 0.012 & 0.371 $\pm$ 0.018 & 0.366 $\pm$ 0.024 &	0.377 $\pm$ 0.016 \\
GloVe	    & 0.410 $\pm$ 0.007 & 0.381 $\pm$ 0.013 & 0.393 $\pm$ 0.016 &	\textbf{0.395 $\pm$ 0.010} \\

 \hline
 \multicolumn{5}{c}{MSPIN-SD} \\
WE	        & 0.120 $\pm$ 0.047 &	0.148 $\pm$ 0.023	& 0.084 $\pm$ 0.024 &	0.105 $\pm$ 0.026 \\
Word2Vec	& 0.138 $\pm$ 0.031 &	0.108 $\pm$ 0.024	& 0.101 $\pm$ 0.024 &	0.116 $\pm$ 0.017 \\
GloVe	    & 0.147 $\pm$ 0.043 &	0.141 $\pm$ 0.019	& 0.098 $\pm$ 0.017 &	\textbf{0.128 $\pm$ 0.015} \\
 \hline
 \multicolumn{5}{c}{IEMOCAP-LOSO} \\
WE	        & 0.376 $\pm$ 0.008 &	0.359 $\pm$ 0.018 & 0.370 $\pm$	0.020 & 0.368 $\pm$ 0.013 \\
Word2Vec	& 0.375 $\pm$ 0.058 &	0.357 $\pm$ 0.058 & 0.365 $\pm$	0.065 & 0.366 $\pm$ 0.059 \\
GloVe	    & 0.405 $\pm$ 0.009 &	0.382 $\pm$ 0.020 & 0.378 $\pm$	0.021 & \textbf{0.389 $\pm$ 0.014} \\
 \hline
 \multicolumn{5}{c}{MSPIN-LOSO} \\
WE	        & 0.076 $\pm$ 0.013 &	0.196 $\pm$ 0.011 & 0.136 $\pm$	0.015 &	0.136 $\pm$ 0.009 \\
Word2Vec	& 0.162 $\pm$ 0.008 &	0.202 $\pm$ 0.005 & 0.147 $\pm$	0.003 &	\textbf{0.170 $\pm$ 0.000} \\
GloVe	    & 0.192 $\pm$ 0.004 &	0.189 $\pm$ 0.007 & 0.129 $\pm$	0.004 &	\textbf{0.170 $\pm$ 0.003} \\
 \hline
 \end{tabular}
\end{table} 

To find the optimal parameter values for $\alpha$ and $\beta$, a linear search
was performed on the scale [0.0, 1.0] with a step of 0.1. Using this
conventional technique, we found four sets of optimal parameters for the
acoustic and text networks. Note that while only the improvised and natural
scenarios (MSP-I+N) were used to find the optimal text-network parameters for
the MSP-IMPROV dataset, the whole dataset was used to find the optimal
acoustic-network parameters. Table \ref{tab:optim_params} lists the optimal
parameter values for $\alpha$ and $\beta$.

\begin{table}[htpb]
 \centering
 \caption{Optimal parameters for multitask learning.}
 \label{tab:optim_params}
 \begin{tabular}{l l c c}
 \hline
 Dataset & Modality & $\alpha$ & $\beta$ \\
 \hline
 IEMOCAP & acoustic & 0.1 & 0.5 \\
 & text & 0.7 & 0.2 \\
 MSP-IMPROV & acoustic & 0.3 & 0.6 \\
 & text & 0.1 & 0.6 \\
 \hline
 \end{tabular}
\end{table} 

To summarize the single-modality results, average CCC scores from three emotion
dimensions can be used to justify which features perform better among others.
The results show that HSF2 was the most useful of the acoustic feature sets (in
two of four datasets), while the word embedding (WE) with pretrained GloVe
embedding was the most useful of the text feature sets. The performance of
dimensional emotion recognition in the speaker-independent (LOSO) case was
lower than in the speaker-dependent (SD) case, as predicted. Note that both
acoustic and text emotion networks used a fixed seed number to achieve the same
result for each run; however, the text network resulted in different scores. Hence,
standard deviations were given to measure fluctuation in 20 runs.

\subsection{Results from SVM-based fusion}
\label{subsect:svm_result}
The main proposal of this research is the late-fusion approach combining the
results from acoustic and text networks for dimensional emotion recognition.
This subsection presents the results of the late-fusion approach, including the
obtained performances, comparison with the single-modality results, which pairs
of acoustic-text results performed better, and our overall findings.

For each dataset (IEMOCAP-SD, MSPIN-SD, IEMOCAP-LOSO, MSPIN-LOSO), nine
combinations of acoustic-text result pairs could be fed to the SVM system.
Tables \ref{tab:svm-iemocap-sd}, \ref{tab:svm-mspin-sd},
\ref{tab:svm-iemocap-loso}, and \ref{tab:svm-mspin-loso} list the respective CCC
results for these datasets. Generally, our proposed two-stage dimensional
emotion recognition improved the CCC score from single-modality emotion
recognition. The pair of results from HSF2 (acoustic) and Word2Vec (text) gave
the highest CCC score on speaker-dependent scenarios.

On the speaker-independent IEMOCAP dataset (IEMOCAP-LOSO), the result from the
pair of HSF2 and GloVe gave the highest CCC score. This result linearly
correlated with the single-modality results for that dataset, in which HSF2
obtained the highest CCC score among the acoustic features, and GloVe was the
best among the text features. On the four datasets, the results from HSF2
obtained the highest CCC score for two out of four datasets, while GloVe
obtained the highest CCC score for all four datasets. Hence, we conclude that
the highest result from a single modality, when paired with the highest result
from another modality, will achieve the highest performance among possible
pairs.

\begin{table}[htpb]
 \centering
 \caption{CCC score results after late fusion using an SVM on the IEMOCAP-SD test set.}
 \label{tab:svm-iemocap-sd}
 \begin{tabular}{l c c c c}
 \hline
 Inputs & V & A & D & Mean \\
 \hline
 LLD	+ WE	    & 0.520 & 0.602 & 0.519 & 0.547 \\
 LLD	+ Word2Vec	& 0.552 & 0.613 & 0.524 & 0.563 \\
 LLD	+ GloVe	    & 0.546 & 0.606 & 0.520 & 0.557 \\
 HSF1	+ WE	    & 0.578 & 0.575 & 0.490 & 0.548 \\
 HSF1	+ Word2Vec	& 0.599 & 0.590 & 0.491 & 0.560 \\
 HSF1	+ GloVe	    & 0.595 & 0.582 & 0.495 & 0.557 \\
 HSF2	+ WE	    & 0.598 & 0.591 & 0.502 & 0.564 \\
 HSF2	+ Word2Vec	& 0.595 & 0.601 & 0.499 & \textbf{0.565} \\
 HSF2	+ GloVe	    & 0.598 & 0.591 & 0.502 & 0.564 \\ 
 \hline
 \end{tabular}
\end{table} 

\begin{table}[htpb]
 \centering
 \caption{CCC score results after late fusion using an SVM on the MSPIN-SD dataset.}
 \label{tab:svm-mspin-sd}
 \begin{tabular}{l c c c c}
 \hline
 Inputs & V & A & D & Mean \\
 \hline
LLD	    + WE	    & 0.344 & 0.591 & 0.447 & 0.461 \\
LLD	    + Word2Vec	& 0.326 & 0.586 & 0.439 & 0.450 \\
LLD	    + GloVe	    & 0.344 & 0.585 & 0.439 & 0.456 \\
HSF1	+ WE	    & 0.461 & 0.637 & 0.517 & 0.538 \\
HSF1	+ Word2Vec	& 0.464 & 0.634 & 0.518 & 0.539 \\
HSF1	+ GloVe	    & 0.466 & 0.630 & 0.510 & 0.535 \\
HSF2	+ WE	    & 0.475 & 0.640 & 0.522 & 0.546 \\
HSF2	+ Word2Vec	& 0.486 & 0.641 & 0.524 & \textbf{0.550} \\
HSF2	+ GloVe	    & 0.485 & 0.638 & 0.523 & 0.549 \\
 \hline
 \end{tabular}
\end{table} 

\begin{table}[htpb]
 \centering
 \caption{CCC score results after late fusion using an SVM on the IEMOCAP-LOSO test set.}
 \label{tab:svm-iemocap-loso}
 \begin{tabular}{l c c c c}
 \hline 
 Inputs & V & A & D & Mean \\
 \hline
LLD	    + WE	    & 0.537 & 0.583 & 0.431 & 0.517 \\
LLD	    + Word2Vec	& 0.528 & 0.580 & 0.421 & 0.510 \\
LLD	    + GloVe	    & 0.539 & 0.587 & 0.430 & 0.518 \\
HSF1	+ WE	    & 0.565 & 0.565 & 0.453 & 0.528 \\
HSF1	+ Word2Vec	& 0.536 & 0.559 & 0.434 & 0.510 \\
HSF1	+ GloVe	    & 0.559 & 0.570 & 0.452 & 0.527 \\
HSF2	+ WE	    & 0.524 & 0.566 & 0.452 & 0.514 \\
HSF2	+ Word2Vec	& 0.531 & 0.571 & 0.445 & 0.516 \\
HSF2	+ GloVe	    & 0.553 & 0.579 & 0.465 & \textbf{0.532} \\
 \hline
\end{tabular}
\end{table} 

\begin{table}[htpb]
 \centering
 \caption{CCC score results after late fusion using an SVM on the MSPIN-LOSO test set.}
 \label{tab:svm-mspin-loso}
 \begin{tabular}{l c c c c}
 \hline 
 Inputs & V & A & D & Mean \\
 \hline
 LLD	+ WE	    & 0.204 & 0.485	& 0.387 & 0.358 \\
 LLD	+ Word2Vec	& 0.267 & 0.487	& 0.386 & 0.380 \\
 LLD	+ GloVe	    & 0.269 & 0.482	& 0.375 & 0.376 \\
 HSF1	+ WE	    & 0.224 & 0.565	& 0.410 & 0.400 \\
 HSF1	+ Word2Vec	& 0.286 & 0.558	& 0.411 & 0.418 \\
 HSF1	+ GloVe	    & 0.282 & 0.555	& 0.409 & 0.415 \\
 HSF2	+ WE	    & 0.232 & 0.566	& 0.421 & 0.406 \\
 HSF2	+ Word2Vec	& 0.287 & 0.562	& 0.411 & 0.420 \\
 HSF2	+ GloVe	    & 0.291 & 0.570	& 0.405 & \textbf{0.422} \\ 
 \hline
\end{tabular}
\end{table} 

To evaluate the improvement obtained by SVM-based late fusion, the average CCC
scores again can be used as a single metric. The rightmost column in Tables
\ref{tab:svm-iemocap-sd}, \ref{tab:svm-mspin-sd}, \ref{tab:svm-iemocap-loso},
and \ref{tab:svm-mspin-loso} shows the average CCC scores obtained from the
nine pairs of acoustic and text results on the four different datasets.
Comparing these bimodal results to unimodal results (Table \ref{tab:ser-test}
and \ref{tab:ter-test}) shows the difference. All results from SVM improved
unimodal results. In speaker-independent (LOSO) results (which are more
appropriate for real-life analysis), the scores resulted by pairs of HSF with
any word vector obtain remarkable improvements, particularly in the MSPIN-LOSO
dataset.  For any other pair involving LLDs, the obtained score was also lower
as compared to other pairs.  Considering all low scores involved LLD results,
improving the performance of dimensional emotion recognition by using LLDs is
more complicated than by using HSF1 and HSF2, apart from the larger feature
size and the longer training time. The large network size created by an LLD
input as a result of its much bigger feature dimension (e.g., 3409 $\times$ 23
on IEMOCAP) did not help either the single-modality or late-fusion performance.
In contrast, the small sizes of the functional features (HSF1 and HSF2) enabled
better performance on a single modality, which led to better performance for
the late-fusion score. To obtain functional features, however, a set of LLD
features must be obtained first. This problem is a challenging future research
direction, especially for implementing dimensional emotion recognition with
real-time processing. 

Aside from the fact that a speaker-independent dataset is usually more
difficult than a speaker-dependent dataset, the low score on MSPIN-LOSO was due
to its low scores on a single modality. In other words, lower pair performance
from a single modality will result in low performance in late fusion. In
particular, these low results derive from low CCC scores from the text modality
(Table \ref{tab:ter-test}). The average CCC score for the text modality on the
MSPIN-LOSO dataset was less than 0.16, compared to an average score higher than
0.34 for the acoustic modality. All nine pairs in late-fusion approaches
improved on the single-modality results because of the two-stage DNN and SVM 
regression analysis. Thus, out of 36 trials (9 pairs $\times$ 4 datasets), our
proposed two-stage dimensional emotion recognition outperforms any single
modality result (used in a pair).

The low score on MSPIN for the text modality can be tracked to the origin of
the dataset; that is, there may have been a number of sentences semantically
identical to the target sentences in the dataset we used. Although we chose
sentences only from the improvised dialogues (minus the target sentence) and
from the natural interactions (those sentences produced by the experimenters
and subjects during the breaks), some of the sentences in this corpus were
semantically identical to that of the target sentences in the
``Target-Improvised" data set.  This was confirmed retroactively by manually
checking the provided transcription and our automatic transcription.  Given the
nature of the elicitation task in a dialogue framework, this is not surprising.
A similarly low result for the text modality on this MSPIN dataset was also
shown in \cite{Zhang2019}. In general, compared to the IEMOCAP dataset, the
MSPIN dataset suffers from low accuracy in recognizing the valence category by
using acoustic and lexical properties. Interestingly, however, those authors
also did not show improvement on the IEMOCAP scripted dataset, another
text-based session in which lexical/text features do not contribute
significantly.

\begin{figure}[htpb]
\centering
\includegraphics[width=\textwidth]{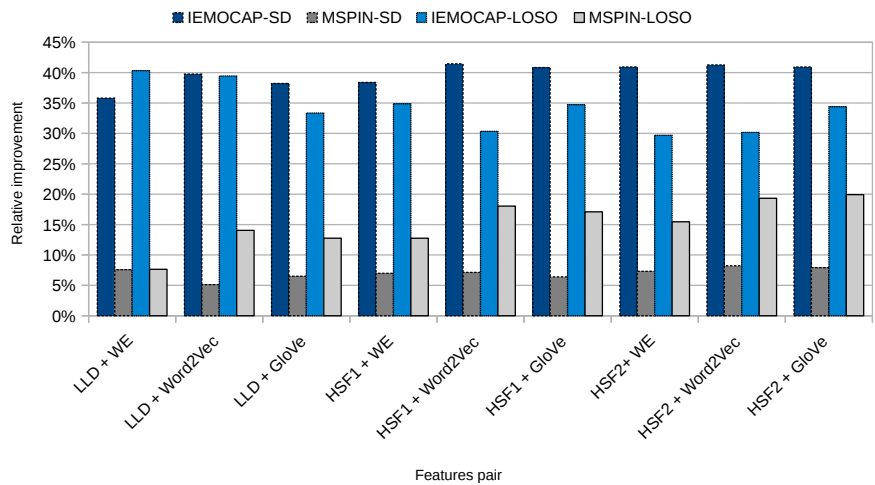}
\caption{Relative improvement in average CCC scores from late fusion using an 
SVM as compared to the highest average CCC scores from a single modality.}
\label{fig:relative-improvement}
\end{figure}

To measure the improvement by our proposed two-stage late fusion, we calculated
the relative improvement obtained by late fusion from the highest CCC scores for
a single modality. For example, the pair of LLD + WE used the results from the
LLDs in the acoustic network and the WE in the text network. We compared the
result for LLD + WE with that for the LLDs, as it had a higher score than the WE
did. Figure \ref{fig:relative-improvement} thus shows the relative improvement
for all nine pairs. All of 36 trials showed improvements ranging from
5.11\% to 40.32\%. Table \ref{tab:relative-improvement} lists the statistics for
the obtained relative improvement. Our results show higher relative accuracy
improvement as compared to those obtained by \cite{Zhang2019} for valence
prediction, which ranged from 6\% to 9\%. Nevertheless, their multistage fusion
method also showed benefits over the multimodal and single-modality approaches.
These findings confirm the benefits of using bimodal/multimodal fusion instead
of single-modality processing for valence, arousal, and dominance prediction.

\begin{table}[htpb]
 \centering
 \caption{Statistics of relative improvement by late fusion using an SVM as
 compared to the highest scores for a single modality across datasets; the
 scores were extracted from the data shown in Figure
 \ref{fig:relative-improvement}.}
 \label{tab:relative-improvement}
 \begin{tabular}{l c c c c}
 \hline
 Statistic & IEMOCAP-SD & MSPIN-SD & IEMOCAP-LOSO & MSPIN-LOSO \\
 \hline
 Average& 39.73\%	& 7.01\%	& 34.15\%	& 15.23\% \\
 Max	& 41.45\%	& 8.22\%	& 40.32\%	& 19.93\% \\
 Min	& 35.80\%	& 5.11\%	& 29.69\%	& 7.64\% \\
 Std	& 1.90\%	& 0.93\%	& 3.84\%	& 3.90\% \\
 \hline
 \end{tabular}
\end{table} 

\subsection{Speaker-dependent vs. speaker-independent text emotion recognition}
\label{subsect:sd_loso}
While speech-based emotion recognition is performed with a fixed random seed to
generate the same result for each run, text-based emotion recognition results
in different scores for each run. The different results on text emotion
recognition probably were caused by the initiation of weightings on embedding
layers.  In this case, statistical tests can be performed on text emotion
results  
to observe the difference between speaker-dependent and speaker-independent 
scenario. In contrast, statistical tests cannot be performed between
acoustic results and bimodal acoustic-text results due to differences in
the data (deterministic vs. non-deterministic). 

Table \ref{tab:test_sd_loso} shows if there is a significant difference between
speaker-dependent and speaker-independent results on the same feature set. We
set $p-$value = 0.05 with a two-tail paired t-test between mean scores of
speaker-dependent and speaker-independent results. This paired t-test was based
on the assumption that there are no outliers (after pre-processing) and two
different inputs are fed into the same system. Only one result from text
emotion recognition shows no significant difference on IEMOCAP dataset while
all results on MSPIN dataset show a significant difference between
speaker-dependent and speaker-independent results. This result reveals a
tendency for a difference in evaluating speaker-dependent and
speaker-independent data. The results from speaker-dependent data were
different from those of speaker-independent data. In other words, results
from speaker-dependent data cannot be used to justify speaker-independent or
whole data.

\begin{table}
    \centering
    \caption{Significant difference between speaker-dependent and speaker-independent scenario on the same text feature set; statistical tests were performed using two-tail paired $t-$test with $p-$value = 0.005.}
    \label{tab:test_sd_loso}
    \begin{tabular}{l c c}
        \hline
        Feature     &   IEMOCAP     & MSPIN \\
        \hline
        WE          &   Yes         & Yes   \\
        Word2Vec    &   No          & Yes   \\
        GloVe       &   Yes         & Yes   \\
    \hline
    \end{tabular}
\end{table}

\subsection{Effect of removing target sentence from MSPIN dataset}
Since the goal of this research is to evaluate the contribution of both
acoustic and linguistic information in affective expressions, it is necessary
to have sentences in the dataset that are free from any stimuli control.
However, the original MSP-IMPROV dataset contains 20 ``target" sentences; a
sentence with the same linguistic content but produced with different
emotions. These parts of MSP-IMPROV dataset are irrelevant to this study;
hence, we remove it from the dataset (\textit{Target - improvised} and
\textit{Target - read} parts).  However, we found that the results show low CCC
scores (Table \ref{tab:ter-test}, particularly on valence) indicating 
influence from target sentences. These results may be explained, as mentioned in
section \ref{subsect:svm_result}, that some utterances in the data
analyzed in this study also inadvertently included sentences semantically the
same as those in the improvised target sentences. 

\subsection{Final remarks}

We tried to perform a benchmark between our results and others on the same
datasets, scenarios, and metrics. Unfortunately, to the best of our knowledge,
the only reference is the one reported by \cite{Atmaja2020d}, which reports an
early fusion method on IEMOCAP dataset. We improve the average CCC score from
$0.508$ to $0.532$. This higher result suggests that late fusion is better than
early fusion in modeling how humans fuse multimodal information, which is in
line with neuropsychological research. This late-fusion approach can be
embedded with current speech technology, i.e., ASR, in which the text output
can be processed to weigh emotion prediction from acoustic features.

\cite{Abdelwahab2018} used MSP-Podcast \citep{Lotfian2019} as a target corpora,
which is not available for the public yet, and IEMOCAP with MSP-IMPROV as
a source corpus to implement their DANN for cross-corpus speech emotion
recognition.  Although the goal is different, we observed similar patterns
between theirs and our acoustic-only speech emotion recognition. First, we
observed that the order of highest to lowest CCC scores is arousal, dominance,
and valence. This pattern is also consistent when IEMOCAP is mixed with
MSP-IMPROV as reported by \cite{parthasarathy2017jointly} (in Table 2). Second,
we observed that the CCC scores obtained in IEMOCAP are higher than those
obtained in MSP-IMPROV; we believe that this lower score in MSP-IMPROV was due
to the smaller size of the dataset. 

Along with our SVM architecture, we also explored the parameters $C$ and
$\gamma$, because both parameters are important for an RBF-kernel-based SVM
architecture \citep{scikit-learn}. Linear search was used in the ranges of
[$10^{-2}, 1, 10^2, 2 \times 10^2, 3 \times 10^2$] for $C$ and [$10^{-2},
10^{-1}, 1, 10, 10^2$] for $\gamma$ with a fixed value of $\epsilon$, i.e.,
0.01. The best parameter values were $C=200$ and $\gamma=0.1$. The repository
includes the detailed implementation of the SVM architecture.

Per our stated objective, we applied two-stage processing by using DNNs and an
SVM for dimensional emotion recognition from acoustic and text features on four
different datasets. We found that the combination of mean + std + silence from
the acoustic features and word embeddings weighted by pretrained GloVe
embeddings achieved the highest result among the nine pairs of acoustic-text
results from DNNs trained with multitask learning. When the performance in
obtaining one input to the SVM is very low, the resulting relative improvement
due to the SVM is also low. For instance, the lowest improvement on MSPIN-LOSO
was from LLD + WE features, in which WE obtained a low score
($\overline{CCC}=0.136$) on text network. This phenomenon suggests a
challenging future research direction for dealing with minimal linguistic
information in the fusion strategy. One strategy applied in this research was
to use a pretrained GloVe embedding on text features with HSF2 on acoustic
features, which improved the $\overline{CCC}$ score from 0.358 (relative
improvement = 7.64\%) to 0.422 (relative improvement = 19.93\%). Other
strategies should also be proposed, such as how to handle the data
differently when the linguistically identical sentences elicit different
emotions (i.e., the whole MSP-IMPROV dataset). In contrast, the currently
evaluated word embeddings treat the same words to have the same
representations, even when it conveys different emotions.

\section{Conclusions}
In conclusion, we summarize several findings. First, we found a linear
correlation between the single-modality and late-fusion methods in dimensional
emotion recognition. The best results from each modality, when they were paired,
gave the best fusion result. In the same way, the worst results obtained from
each network, when they were paired, gave the worst fusion results for bimodal
emotion recognition. This finding differs from that reported in
\cite{Atmaja2019b}, which used an early-fusion approach for categorical emotion
recognition. In their work, the best pair differs from the best methods in
single modalities.

Second, text features strongly influenced the score of dimensional SER on the
valence dimension, while acoustic features strongly influenced arousal and
dominance scores. Accordingly, the proposed two-stage processing can take
advantage of text features that are commonly used in predicting sentiment
(valence) for the dimensional emotion recognition task. The proposed fusion
method improves all three emotion dimensions without attenuating the
performance of any dimension.  That is, the proposed method elevates the scores
for valence, arousal, and dominance subsequently from the highest to the lowest
gain.

Third, the combination of input pairs does not matter in the proposed fusion
method, as indicated by the low deviation in relative improvement across the
nine possible input pairs. What does matter is the performance of the input in
the DNN stage. If the performance of a feature set in the DNN stage is low
($\overline{CCC} \leq 0.2$), it will also result in low performance when paired
with another low-performance input in the SVM stage.

Finally, this bimodal approach can be extended to a multimodal approach. Both
acoustic and text features can be combined with visual and motion-capture
measurements that have advantages in specific emotion dimensions (liking or
naturalness). The results can be benchmarked with current results to observe
such improvements by adding more modalities. The SVM stage itself can be
performed many times to obtain such improvements. These broad research
directions are open challenges for researchers in human-computer interaction.

\bibliography{CSL2020}

\end{document}